

\input harvmac
\input epsf
\ifx\epsfbox\UnDeFiNeD\message{(NO epsf.tex, FIGURES WILL BE
IGNORED)}
\def\figin#1{\vskip2in}
\else\message{(FIGURES WILL BE INCLUDED)}\def\figin#1{#1}\fi
\def\ifig#1#2#3{\xdef#1{fig.~\the\figno}
\goodbreak\midinsert\figin{\centerline{#3}}%
\smallskip\centerline{\vbox{\baselineskip12pt
\advance\hsize by -1truein\noindent\footnotefont{\bf
Fig.~\the\figno:} #2}}
\bigskip\endinsert\global\advance\figno by1}
\overfullrule=0pt
\parindent 25pt
\tolerance=10000


\def\RR{{$R\otimes R$}}
\def\ZZ {{\bf Z}}

\def\Sl#1{Sl(#1,\ZZ)}
\def\G(#1){\Gamma(#1)}

\def\C|#1{{\cal #1}}
\def\(#1#2){(\zeta_#1\cdot  \zeta_#2)}

\def\Imagi{\Im{\rm m}}

\def\half{{\textstyle {1 \over 2}}}
\def\calT{{\cal T}}
\def\Rfour{t_8t_8R^4}

\def\calD{{\cal D}}
\def\calF{{\cal F}}
\def\calP{{\cal P}}

\def\half{{\textstyle {1 \over 2}}}

\def\hK{\hat K}

\def\calR{{\cal R}}
\def\L{L}

\def\xxx#1 {{hep-th/#1}}
\def\lr{\lref}
\def\npb#1(#2)#3 { Nucl. Phys. {\bf B#1} (#2) #3 }
\def\rep#1(#2)#3 { Phys. Rept.{\bf #1} (#2) #3 }
\def\plb#1(#2)#3{Phys. Lett. {\bf #1B} (#2) #3}
\def\prl#1(#2)#3{Phys. Rev. Lett.{\bf #1} (#2) #3}
\def\physrev#1(#2)#3{Phys. Rev. {\bf D#1} (#2) #3}
\def\ap#1(#2)#3{Ann. Phys. {\bf #1} (#2) #3}
\def\rmp#1(#2)#3{Rev. Mod. Phys. {\bf #1} (#2) #3}
\def\cmp#1(#2)#3{Comm. Math. Phys. {\bf #1} (#2) #3}
\def\mpl#1(#2)#3{Mod. Phys. Lett. {\bf #1} (#2) #3}
\def\ijmp#1(#2)#3{Int. J. Mod. Phys. {\bf A#1} (#2) #3}

\lr\rfGreenKwonVanhove{M.B. Green, H. Kwon and P. Vanhove, {\sl Two-loops in Eleven Dimensions}, DAMTP-99-122, hep-th/9910055.}
\lr\rfGreenCargese{M.B. Green, {\sl Connections between M-theory and
    superstrings}, Proceedings of the 1997 Advanced Study Institute on
    Strings, Branes and Dualities, Cargese, France Nucl.Phys.Proc.Suppl. 68
    (1998) 242,  hep-th/9712195.} 
\lr\rfEulersums{O.M.~Ogreid and P.~Osland, {\sl Summing one- and
    two-dimensional series related to the Euler series\/},
    J.Comput.Appl.Math. {\bf 98} (1998) 245; hep-th/9801168.}
\lr\rfRussoTseytlin{J.G. Russo and A.A. Tseytlin, {\sl One-loop four-graviton
    amplitude in eleven-dimensional supergravity}, Nucl.Phys. {\bf B508} (1997)
  245; hep-th/9707134.}
\lr\rfTerrasBook{A. Terras, {\sl Harmonic analysis on symmetric spaces and
    applications I \& II}, Springer-Verlag 1985.}
\lr\rfSakaiTanii{N.~Sakai and Y.~Tanii, {\sl One-loop amplitudes and effective
    action in superstring theories}, Nucl.Phys. {\bf B 287} (1987) 457.}
\lr\rfDhokerPhongAnalytic{E. D'Hoker and D. H. Phong, {\sl The Box Graph In
    Superstring Theory}, Nucl. Phys. {\bf B440} (1995) 24;
    hep-th/9410152.;\hfill\break
E. D'Hoker and D.H. Phong, {\sl Momentum Analyticity and Finiteness of the
    1-Loop Superstring Amplitude}, Phys.Rev.Lett. {\bf 70} (1993) 3682;
    hep-th/9302003.}
\lr\rfDhokerPhongRevue{E. D'Hoker and D.H. Phong, {\sl The geometry of string
    perturbation theory} , Rev. Mod. Phys. {\bf 60} (1988) 917.}
\lr\rfGreenSchwarzWitten{M.B. Green, J.H. Schwarz and E. Witten, {\sl
    Superstring theory}, Cambridge University Press.}
\lr\rfLercheNilssonSchellekensAnomaly{W. Lerche, B.E.W. Nilsson and
    A. Schellekens, {\sl Heterotic String loop Calculation of the Anomaly
    Cancelling Term}, Nucl.Phys. {\bf B289} (1987) 609.}
\lr\rfLercheNilssonSchellekensWarner{W. Lerche, B.E.W. Nilsson,
    A.N. Schellekens and N.P. Warner, {\sl Anomaly Cancelling Terms from the
    Elliptic Genus}, Nucl.Phys. {\bf B299} (1988) 91.}
\lr\rfGreenGutperleVanhove{M.B. Green, M. Gutperle and P. Vanhove, {\sl One-loop in Eleven Dimensions}, Phys.Lett. {\bf B409} (1997) 177; hep-th/9706175.}
\lr\rfGreenSchwarzone{M.B. Green and J.H. Schwarz, {\sl Supersymmetrical Dual String Theory. 2. Vertices and Trees.}, Nucl.Phys. {\bf B198} (1982) 252.}
\lr\rfGreenSchwarztwo{M.B. Green and J.H. Schwarz, {\sl Supersymmetrical Dual String Theory. 3. Loops and Renormalization.}, Nucl.Phys. {\bf B198} (1982) 441.}
\lr\rfpolchtasi{J.  Polchinski, {\sl TASI lectures on D-branes},
    hep-th/9611050.}
\lr\rfRusso{J.G.  Russo, {\sl Construction of $SL(2,Z)$ invariant
amplitudes in type IIB superstring theory}, hep-th/9802090; Nucl. Phys. {\bf
B535} (1998) 116.}
\lr\rfGreenMoore{M.B. Green and G.  Moore, unpublished.}
\lr\rfGreenGutperleVanhove{M.B. Green, M. Gutperle and P. Vanhove, {\sl
    One-loop in Eleven Dimensions}, Phys.Lett. {\bf B409} (1997) 177;
    hep-th/9706175.} 
\lr\rfGreenVanhoveMtheory{M.B. Green and P. Vanhove, {\sl D-instantons,
    Strings and M-theory}, Phys.Lett. {\bf 408B} (1997) 122; hep-th/9704145.}
\lr\rfGreenGutperle{M.B. Green and M. Gutperle,
 {\sl Effects of D-instantons},
Nucl.\ Phys.\ {\bf B498}, 195 (1997);  hep-th/9701093.}
\lr\rfGreenSethi{M.B.~Green and S.~Sethi, {\sl
Supersymmetry constraints on type IIB supergravity},
Phys.\ Rev.\ {\bf D59}, 046006 (1999);  hep-th/9808061.}
\lr\rfBoris{B. Pioline, {\sl A note on non-perturbative $R^4$ couplings},
    Phys.Lett. {\bf B431} (1998) 73, hep-th/9804023.}
\lr\rfHerveAlex{H. Partouche and A. Kehagias, {\sl On the exact quartic
    effective action for the type IIB superstring}, Phys.Lett. {\bf B422}
    (1998) 109, hep-th/9710023; {\sl D-Instanton Corrections as (p,q)-String
    Effects and Non-Renormalization Theorems}, Int.J.Mod.Phys. {\bf A13}
    (1998) 5075, hep-th/9712164.}

\noblackbox \baselineskip 14pt plus 2pt minus 2pt
\Title{\vbox{\baselineskip12pt \hbox{hep-th/9910056}
\hbox{DAMTP-99-124} \hbox{CERN-TH/99-200} \hbox{SPHT-T99/071 }}}
{\vbox{ \centerline{The low energy expansion of the one-loop}
\smallskip
\centerline{ type II superstring amplitude} }}

\centerline{Michael B. Green and Pierre Vanhove}
\medskip
\centerline{DAMTP, Silver Street, Cambridge CB3 9EW, UK}
\centerline{\tt m.b.green@damtp.cam.ac.uk,
p.vanhove@damtp.cam.ac.uk}
\bigskip

\medskip
\centerline{{\bf Abstract}}

The one-loop four-graviton amplitude in either of the type II superstring
theories is expanded in powers of the external momenta up to and
including  terms of order $s^4  \, \log s\, \calR^4$,
where $\calR^4$ denotes a specific contraction of
four linearized Weyl tensors and $s$ is a Mandelstam invariant.
Terms in this series are obtained
by integrating  powers of the two-dimensional scalar field
theory propagator over   the toroidal world-sheet as well as the
moduli of the torus.  The values of these coefficients match
expectations based on duality relations between string theory and
eleven-dimensional supergravity.

\noblackbox \baselineskip 14pt plus 2pt minus 2pt

\Date{PACS: 04.50.+h}

\newsec{Introduction}

The wealth of duality symmetries relating different
parameterizations of nonperturbative string theory, or M theory,
provide severe constraints on its structure. One
striking manifestation of this is the relationship between  the low
energy expansion of the type II string theory action and one-loop
effects in compactified eleven-dimensional supergravity
\refs{\rfGreenGutperleVanhove}.  Although the systematics of this relationship becomes very murky at higher loops,
the leading behaviour of the
two-loop contribution of the eleven-dimensional theory is
amenable to a detailed
analysis (see the companion paper \refs{\rfGreenKwonVanhove}).

This detailed  comparison between string theory and eleven-dimensional
supergravity requires, among other things, detailed
knowledge of the low energy expansion of
the effective action  of the
type IIA and type IIB superstring perturbation theories.
Surprisingly, this has scarcely been considered in the literature
beyond the most elementary tree-level terms.
 In this paper we will obtain  terms in the effective action that
arise from the momentum expansion of the
one-loop type II superstring theory  contribution to the
four graviton amplitude. Since the four-graviton tree and one-loop
amplitudes in the   IIA and IIB
 theories  are equal  we need not distinguish between the two theories in the
following\foot{The two type II string perturbation theories are equal up
to and  including two loops
\refs{\rfGreenKwonVanhove}}.

 The tree-level amplitude for the scattering of
four gravitons with polarization tensors $\zeta^{(r)}_{\mu \nu}$ and
momenta $k_r^\mu $ ($r=1,2,3,4$, $\mu =0,1,\dots,9$
  and $k_r^2 =0$) has the very simple form \refs{\rfGreenSchwarztwo,\rfDhokerPhongRevue}
\eqn\eTree{A_4^{tree} = - \hK\, \kappa_{10}^2\, e^{-2\phi}T,}
where $\phi$ is the constant dilaton field so that $g=\kappa_{10}^{-1}e^\phi$
is the string coupling and
\eqn\etdef{T = {64\over{\alpha'}^3 stu}
{\Gamma\left(1-{\alpha'\over 4} s \right)\Gamma\left(1-{\alpha'\over 4} t \right)
\Gamma \left(1-{\alpha'\over 4} u \right)\over
  \Gamma\left (1 +{\alpha'\over 4} s \right)\Gamma\left(1+{\alpha'\over 4} t\right)
  \Gamma\left (1+{\alpha'\over 4} u \right)},}
where the Mandelstam invariants are defined by $s=-(k_1+k_2)^2$,
$t=-(k_1+k_4)^2$ and $u=-(k_1+k_3)^2$. The overall kinematic factor  $\hK$
 is given by
 \eqn\useam{ \hK=
t^{\mu _1\dots \mu _8} t^{\nu_1\dots \nu_8} \prod_{r=1}^4 \zeta^{(r)}_{\mu _r\nu_r}
k^{(r)}_{\mu _r} k^{(r)}_{\nu_r},}
which is the linearized approximation to
 the standard contraction between four curvature tensors,
\eqn\eTensterm{
\calR^4=\Rfour\equiv t^{\mu  _1\dots \mu  _8} t_{\nu_1\dots \nu_8}
R^{\nu_1\nu_2}_{\mu  _1\mu  _2} \cdots R^{\nu_7\nu_8}_{\mu  _7\mu 
_8}, }
where the tensor $t^{\mu  _1\dots \mu  _8}$
 is defined in \refs{\rfGreenSchwarzone} and in
Appendix~9 of \refs{\rfGreenSchwarzWitten}\foot{This contraction
projects onto the purely traceless components of the curvature,
which constitute the Weyl tensor.}.
The value of the constant $\kappa_{10}$ in \eTree\ is arbitrary since it can
be changed by shifting the dilaton field.  It is convenient
to set it to the value
\eqn\ekapdef{\kappa_{10}^2 = {1\over 2}(2\pi)^7{\alpha'}^4,}
which normalizes the D-string tension to the value $T_{D_1} =
e^{-\phi} T_F$, where $T_F =1/2\pi\alpha'$ is the fundamental
string tension \refs{\rfpolchtasi}.

The one-loop type II superstring four-graviton scattering amplitude in ten
dimensions is also
very simple and is given by \refs{\rfGreenSchwarzone}\foot{There is a
factor of 2 difference in this and a few subsequent formulae compared
to  the first version of this paper.  We are grateful to J. Russo for
pointing  out these  factors.} 
\eqn\eLoopa{ A_4^{one-loop}= {\kappa_{10}^4\over 2^5\, \pi^6\alpha'{}^4}\hat K
  I  = \kappa_{10}^2  2\pi \, \hat K\, I, }
where $I$ is the integral of a
modular function,
\eqn\eLoop{
I= \int_\C|F {d^2\tau\over \tau_2^2} F(\tau,\bar \tau)}
(where $\tau=\tau_1+i\tau_2$ and $d^2\tau \equiv d\tau_1d\tau_2 =
d\tau d\bar\tau/2$) and
$\C|F$ denotes the fundamental domain of $\Sl2$,
\eqn\fundom{ \C|F=\{|\tau_1|\leq \half, |\tau|^2 \geq 1\}.}
The dynamical factor in \eLoop\ is given by an integral over the positions
$\nu^{(i)} = \nu^{(i)}_1 + i \nu^{(i)}_2$ of the four vertex operators on the
torus,
\eqn\eF{\eqalign{ F(\tau,\bar \tau)& =  \int_{\C|T} \prod_{i=1}^3
{d^2\nu^{(i)}\over \tau_2}
\left(\chi_{12}\chi_{34}\right)^{\alpha' s}
\left(\chi_{14}\chi_{23} \right)^{\alpha' t}
\left(\chi_{13}\chi_{24} \right)^{\alpha'u}\cr
 &= \int_{\C|T}
\prod_{i=1}^3 {d^2\nu^{(i)}\over \tau_2} \ e^\calD = \int_{\C|T}
\prod_{i=1}^3 {d^2\nu^{(i)}\over \tau_2} \ \exp(\alpha's\Delta_s
+\alpha' t \Delta_t + \alpha' u\Delta_u),\cr }}
where  $d^2\nu^{(i)} \equiv d\nu^{(i)}_1 d\nu^{(i)}_2$,
$\nu^{(4)}=\tau$,  and
\eqn\ppdef{\calD =\alpha's\Delta_s
+\alpha' t \Delta_t + \alpha' u\Delta_u,}
with
 \eqn\deldefs{\Delta_s =  \ln (\chi_{12}\chi_{34}), \qquad
\Delta_t =  \ln(\chi_{14}\chi_{23}),\qquad  \Delta_u = \ln
(\chi_{13}\chi_{24})}
  and $\ln \chi_{ij}(\nu^{(1)} - \nu^{(j)};\tau)$ is the scalar Green function
between the vertices labelled $i$ and $j$ on the toroidal
world-sheet.   These Green functions are integrated over the
domain $\C|T$ defined by
 \eqn\dimar{
\C|T=\{-{1\over 2}\leq \nu_1 < {1\over 2},\; 0\leq
 \nu_2 < \tau_2\} }
It is understood  that the mass shell condition
\eqn\masscon{s+t+u=0}
is enforced in all expressions which ensures that only
conformally invariant ratios of $\chi_{ij}$'s arise in \eF.  For
example,  substituting  $u=-s-t$ the exponent of \eF\ contains
\eqn\confrat{{\chi_{12}\chi_{34}\over \chi_{13}\chi_{24}}, \qquad
{\chi_{14}\chi_{23} \over \chi_{13}\chi_{24}}.}
 This also ensures that   the integrand is
   modular invariant.  Many of the following formulae will be
    expressed in a symmetric form in  terms of $s$, $t$ and $u$
   even though these variables are related by
   the condition \masscon.
    The relative normalization between the
 two  terms in \eTree\ and \eLoop\ can be  determined by   unitarity
  as in  \refs{\rfSakaiTanii,\rfDhokerPhongRevue}.

  The tree-level string amplitude \eTree\ is sufficiently simple that it is
easily expanded  to all orders in powers of the momentum.
Successive terms in this expansion lead to terms in the effective
action that are polynomials in derivatives acting on $\calR^4$.
The expansion of $T$  begins with the terms
\eqn\begexp{T= {64\over
{\alpha'}^3 stu}  + 2\zeta(3)+ \cdots   .}
Substitution of the first
term in \eTree\
reproduces the tree diagrams of classical ten-dimensional ${\cal N}=2$ supergravity
  which have poles in the
$s$, $t$ and $u$ channels.   The second
term gives the leading correction to the supersymmetric
Einstein--Hilbert theory and determines a term in the effective action
 proportional to
$\calR^4$.   Subsequent terms give information on higher derivative interactions.
 The complete tree-level expansion will be reviewed in section 2.

  The one-loop string amplitude \eLoop\
also has a remarkably simple form --- the overall kinematic factor
 multiplies an integral over the moduli space of
the toroidal world-sheet that is constructed entirely from the scalar
world-sheet propagator. The leading contribution is
proportional to $\calR^4$ but the nonleading terms in the
momentum expansion have not been calculated up to now.
 From general principles we
can anticipate that the momentum expansion of the one-loop
amplitude has the structure,
\eqn\loopexp{\eqalign{
I(s,t) = &   a +  {\alpha'\over 4} I_{nonan\, 1}(s,t,u) + b{{\alpha'}^2 \over
  16} (s^2+t^2+u^2)  \cr
 & + c{{\alpha'}^3\over 64} (s^3+t^3+u^3)  + d {{\alpha'}^4\over 256} (s^4
 +t^4+u^4 ) + {{\alpha'}^4\over 256} I_{nonan\, 2}(s,t,u) +  \cdots
\cr = & I_{an}(s,t,u) + I_{nonan}(s,t,u), \cr}}
 where  $a,b,c,d,\dots$ are constant  coefficients.
 Up to this order the polynomials in the
 Mandelstam invariants are the unique expressions that are $s,t,u$ symmetric.
 These  make up the analytic part of the amplitude,
$I_{an}(s,t,u)$, whereas the  non-analytic
threshold terms
\eqn\einon{I_{nonan}(s,t,u)
= {\alpha'\over 4} I_{nonan\, 1}(s,t,u) + {{\alpha'}^4\over 256} I_{nonan\, 2}(s,t,u) + o({\alpha'}^4)
,}
have logarithmic singularities.
 The  presence of such singularities  follows very simply as a consequence
 of perturbative
 unitarity  due to the phase space available for massless
 two-particle intermediate states.  For energies such that $s<4 {\alpha'}^{-1}$
(i.e., below the first massive string threshold) the amplitude
$A^{one-loop}_4(s,t)$ satisfies the unitarity relation
\eqn\discone{
\eqalign{
{\rm Disc}\, A^{one-loop}_4(s,t)={1\over (2\pi)^2} \int& d^{10}p_1 d^{10}p_2
\ A_4^{tree}(k_1,k_2,-p_1,p_2)\,
\left(A_4^{tree}(k_3,k_4,p_1,-p_2)\right)^\dagger\cr
&\times  \delta^{(10)}(p_1+p_2-k_1-k_2)\;\theta(p_1^0)\delta^{(10)}
(p_1^2)\theta(p_2^0)\delta^{(10)}(p_2^2) .
}}
Substituting the lowest-order
(Einstein--Hilbert) tree-level term from \begexp\   into both factors of
$A_4^{tree}$ on the right-hand side of \discone\ leads immediately to the
$I_{nonan\, 1}$ term in \loopexp.  Substitution of the term with coefficient
$\zeta(3)$ from \begexp\ into one of the factors of $A_4^{tree}$ and the
Einstein--Hilbert term into the other leads to the $I_{nonan\, 2}$ term in
\loopexp, which has three extra powers of $\alpha'$.
These terms will be discussed in more detail in sections 3 and 4 (see
also \refs{\rfDhokerPhongAnalytic,\rfRussoTseytlin,\rfGreenCargese}).

 The main purpose of this paper is to evaluate a number of  terms in the
 expansion \loopexp. This exercise involves integrating modular invariant
 combinations products of the scalar field theory propagators $\ln \chi_{ij}$
 over the toroidal world-sheet  as well as integration over the moduli space
 of the torus.  Although the integration of combinations of {\it derivatives} of
 world-sheet scalar propagators has arisen in the literature, for example, in
 connection with the elegant calculation of the elliptic genus
 \refs{\rfLercheNilssonSchellekensWarner}, in order to perform the  integrals
 that arise in this paper
 we will need to use some tricks that that  will be presented in
 section 4. This will allow us to determine all the terms in \loopexp\ up to
 order ${\alpha'}^4I_{nonan\, 2}$ (although  the value of the coefficient $d$
will be left as a quadruple sum).  The values of these coefficients are
 compared in \refs{\rfGreenKwonVanhove} with the values that emerge by
 considering two-loop eleven-dimensional supergravity compactified on a
 two-torus.

\newsec{Overview of the tree amplitude}

 The tree amplitude for the scattering of four gravitons of
momenta $k_1^\mu $, $k_2^\mu $, $k_3^\mu $ and $k_4^\mu $ in either
of the type II superstring theories is given by \eTree\ and \etdef\ where
$T$ can be written as \refs{\rfGreenSchwarzone},
\eqn\eTreeta{ T= {64 \over {\alpha'}^3stu}
\exp\left(\sum_{n=1}^\infty {2 \zeta(2n+1) \over
2n+1}\left({{\alpha'}\over 4}\right)^{2n+1} (s^{2n+1} + t^{2n+1} + u^{2n+1})\right),}
where we have used the
elementary identity $\ln\Gamma(1-z)=\gamma\,z+\sum_{n>1} \zeta(n) z^n/n$.

It is convenient to introduce the notation
 $\sigma_k=(\alpha'/ 4)^k\, (s^k+t^k+u^k)$  ($\sigma_1=0$),
 which satisfies the recursion  relation
\eqn\eRecursion{
\sigma_{3+j}={1\over 2} \sigma_2 \sigma_{j+1} + {1\over 3}
 \sigma_3 \sigma_j, \quad \forall j>0 .}
 The solution of these  conditions can be expressed by the generating function,
\eqn\formser{
\sum_{j=1}^\infty x^j \sigma_j = {x^2\sigma_2+x^3\sigma_3\over
1-{1\over 2}\sigma_2
  x^2- {1\over 3} \sigma_3 x^3}
  =(x^2\sigma_2+x^3\sigma_3) \sum_{k\geq 0} x^k
\left(\sum_{2p+3q=k} {(p+q)!\over p! q!}   \left({\sigma_2\over 2}\right)^p
\left({\sigma_3\over 3}\right)^q\right)}
  Therefore
\eqn\eSigma{
\sigma_k = k   \sum_{2p+3q=k} {(p+q -1)!\over p!\, q!}
\left({\sigma_2\over 2}\right)^p
\left({\sigma_3\over 3}\right)^q.}

Since  ${\alpha'}^3 stu/64=\sigma_3/3$ and every $\sigma_{2n+1}$ is
divisible by   $\sigma_3$, the expansion of the exponential in
\eTree\  can be expressed entirely in terms of polynomials of
$\sigma_2$ and $\sigma_3$,
\eqn\eTreeExp{  T=   {3\over \sigma_3}
+ 2\zeta(3)   +   \zeta(5) \sigma_2 + {2\over
  3} \zeta(3)^2   \sigma_3  +  {1\over 2} \zeta(7) (\sigma_2)^2  +
   {2\over 3}  \zeta(3) \zeta(5)
  \sigma_2\sigma_3
  + \cdots.}
It will be significant for the later discussion of unitarity that
the series of powers of $s$, $t$ and $u$ has gaps of three powers
of the Mandelstam invariants between the first two terms and two
powers between the second and third terms.  Each term translates
into a term in the effective action of the type IIB string theory
which is the linearized version of a number of covariant
derivatives acting on $\calR^4$.  These higher derivative terms
are part of the full duality-invariant effective action for the
type IIB string.

\newsec{Expansion of the one-loop amplitude}

In this section and the next we will
 consider the low energy expansion of the one-loop
integral \eLoop\ in powers of $s$, $t$ and $u$.  Formally, this
involves expanding the integrand $F(\tau,\bar\tau)$ \eF\ in
powers of the scalar Feynman propagator which are then integrated
over the toroidal world-sheet,
\eqn\anloop{
I =  \int_\C|F {d^2\tau\over \tau_2^2} \, F(\tau,\bar \tau)
 =  \sum_{n=0}^\infty
  \int_\C|F {d^2\tau\over \tau_2^2} \int_{\C|T}
\prod_{i=1}^3 {d^2\nu^{(i)}\over \tau_2} \,{1\over n!}  \C|D^n
}
 where the exponent is given by
\eqn\expon{\C|D= \alpha' s\, \ln (\chi_{12}\chi_{34})+ \alpha' t\, \ln(\chi_{14}\chi_{23})+ \alpha' u\, \ln (\chi_{13}\chi_{24}) .}
 This expansion is only formal
since we already know that the amplitude is not analytic at
$s=0$, $t=0$ or $u=0$.  This lack of analyticity is manifested by
divergent coefficients in the series \anloop.  One way of dealing
with this problem would be to consider the expansion in a power
series around a nonzero value of $s$, $t$ and $u$ $\sim
\epsilon$. The terms that are singular in the $\epsilon\to 0$
limit can then be resummed  to give the logarithmic
singularities.

A more straightforward procedure is to evaluate the coefficients
of the derivatives of $I$ in the small $s$, $t$ and $u$ limit.
We will therefore consider the general term,
\eqn\genderiv{\eqalign{&I_{an}^{(m,n)} = \lim_{s,t\to 0}
(I^{(m,n)} - I^{(m,n)}_{nonan}) \equiv  \lim_{s,t\to 0} (4 {\alpha'}^{-1})^{m+n} \partial_s^m\partial_t^n (I - I_{nonan}) \cr
 =& \lim_{s,t\to 0} \left( \int_\C|F {d^2\tau\over \tau_2^2} \int_{\C|T}
  \prod_{i=1}^3 {d^2\nu^{(i)}\over \tau_2} \, (4\Delta_s-4\Delta_u)^m\,
  (4\Delta_t -4\Delta_u)^n \,  \exp(\alpha' s \Delta_s +\alpha' t \Delta_t +
\alpha'u \Delta_u) \right. \cr
 & \left .\qquad - I^{(m,n)}_{nonan}\right),\cr}}
where $\Delta_s$, $\Delta_t$ and
$\Delta_u$ are defined in \deldefs\ and $I_{nonan}^{(m,n)}
=(4{\alpha'}^{-1})^{m+n} \partial_s^m\partial_t^n I_{nonan}$.

Since $I_{nonan}$ has logarithmic branch points the $I_{nonan}^{(m,n)}$
terms are
singular functions of $s$ and $t$ which must be extracted from the
complete expression, \genderiv, before the analytic terms can be
determined. Since the
 nonanalytic  terms originate, via unitarity,  from the logarithmic
normal thresholds due to  on-shell intermediate states we can anticipate
that they arise from the region of moduli
space in which $\tau_2\to \infty$, which is the degeneration
limit of the torus. Our strategy in calculating $I^{(m,n)}$
will therefore be to introduce a cut-off $\L$ at a finite but large
value of $\tau_2$.  The region $\tau_2\le \L$ gives
a finite contribution to $I^{(m,n)}$ which includes $I_{an}^{(m,n)}$
together with a $\L$-dependent term.   In this region the exponential
 factor in the integrand on the right-hand side of \genderiv\ can be
 replaced  by unity.  However,  for
$\tau_2\ge \L$ the exponential factor plays a crucial r{\^o}le
in regulating the integral, resulting in  the terms in
$I_{nonan}^{(m,n)}$ together with another finite $\L$-dependent
piece.
Dependence on $\L$ cancels out in the full expression.
 These nonanalytic terms will be considered in detail in
section 3.3, section 4.3 and the appendix.

Differentiating the analytic terms in \loopexp\ an appropriate number of times
with respect to $s$ and $t$ we see that the coefficients
that will be extracted from \genderiv\ have the form
(up to fourth order),
\eqn\sderivs{\eqalign{ & I_{an}^{(0,0)} =  a , \quad\qquad I_{an}^{(1,0)}=0,
\qquad
    I_{an}^{(2,0)} =  4b = 2I^{(1,1)}_{an} ,\qquad   \cr  &  I_{an}^{(2,1)} =
    -6c , \qquad I_{an}^{(3,0)} = 0 , \qquad I_{an}^{(3,1)} =  24 d =
    I_{an}^{(2,2)} , \qquad I_{an}^{(4,0)}
 = 48 d  ,\cr}}
together with the terms obtained by interchanging $s$ with $t$.
The numerical
values of the  coefficients $a$, $b$ and $c$ will be determined in section
4, although $d$ will be left in the form of a multiple sum that
will not be evaluated.

\subsec{The scalar propagator on a torus}

The exponent, $\calD =\alpha'(s \Delta_s + t \Delta_t + u \Delta_u)$,
in the expression \eF\ is a linear combination of scalar
world-sheet propagators joining the locations of the four vertex
operators.
 The scalar propagator between two complex points, $\nu^{(i)}=
\nu_1^{(i)} + i \nu_2^{(i)}$ and $\nu^{(j)}=\nu_1^{(j)} + i
\nu_2^{(j)}$, on a torus of modulus $\tau$ is the doubly periodic
function of $\nu^{(ij)}=\nu^{(i)} -\nu^{(j)}$  in the domain  $\C|T$
   that has a logarithmic short distance
singularity.   Thus, the  propagator,
\eqn\oneprop{ \C|P(\nu^{(ij)}|\tau)  =
\ln\chi_{ij}(\nu^{(ij)}, \tau),}
satisfying  toroidal boundary
conditions can be written as a sum  over  image propagators as
\eqn\ePropTheta{ \C|P(\nu|\tau)  =
- {1\over 2}\left( \sum_{n,m\in\ZZ} \ln|\nu+m+n\tau| - \sum_{(m,n)\neq(0,0)}
\ln |m+n\tau|\right) + {\pi \nu_2^2\over 2\tau_2},}
where the last term is the zero mode of the Laplacian.  The
propagator  can also be expressed as
\eqn\newv{\eqalign{ \C|P(\nu|\tau)
&=-{1\over 4} \ln\left|\theta_1(\nu|\tau)\over
\theta_1'(0|\tau) \right|^2 +{\pi \nu_2^2\over 2\tau_2}
\cr &={\pi \nu_2^2\over
  2\tau_2}- {1\over4} \ln\left|\sin(\pi \nu)\over \pi \right|^2 -
\sum_{m\geq1} \left({q^m\over 1-q^m} {\sin^2(m\pi \nu)\over m}
  +c.c.\right),
}}
  where $q= \exp(2i\pi\tau)$ and $\theta_1(\nu|\tau)$ is a standard
Jacobi theta function.

  Another representation of
the propagator that we will use is  obtained by Fourier
transforming with respect to $\nu$, which leads to an
expression in terms of the sum over the discretized momentum
$m\tau+n$,
\eqn\eprop{\eqalign{
\C|P(\nu|\tau)=&  {1\over 4\pi}  \sum_{(m,n)\neq(0,0)}
{\tau_2\over |m\tau+n|^2} \exp\left[2\pi i m\left(\nu_1 -
\tau_1{\nu_2\over \tau_2}\right) -2\pi i n {\nu_2\over
\tau_2}\right] + C(\tau,\bar\tau) \cr =&
 {1\over 4\pi} \sum_{(m,n)\neq(0,0)} {\tau_2\over
|m\tau+n|^2} \exp\left[{\pi\over \tau_2}\left(\bar\nu (m\tau+n) -
    \nu(m\bar\tau+n)\right)\right] + C(\tau,\bar\tau).
\cr}} The zero mode is given by
\eqn\zermod{ C(\tau,\bar\tau) =   {1\over
  2}\ln\left|(2\pi)^{1/2}\eta(\tau) \right|^2,}
where $\eta(\tau)$ is the standard Dedekind function.

The combination of propagators that enters the amplitude is one
for which the zero mode, $C$,  cancels out.  This is a crucial
point in considering the modular invariance of the integrand.
  The group $Sl(2,\ZZ)$ is generated by the two elements $T: \tau\to\tau+1,$
$\nu\to\nu$ and $S: \tau\to -1/\tau,$ $\nu\to\nu/\tau$. Under these
transformations the propagator transforms as
\eqn\groupy{\eqalign{
T:&\quad \C|P(\nu|\tau+1)=\C|P(\nu|\tau)\cr
S:&\quad \C|P\left(\nu\over\tau\right|\left. -{1\over\tau}\right) =
\C|P(\nu|\tau)+ {1\over 2}\ln|\tau| \ ,\cr}}
so the propagator has a modular anomaly which comes from the zero
mode, $C$, in \eprop.  However, the sum over propagators in the exponent $\calD$
is modular invariant since the zero modes cancel after using the
on-shell condition, $s+t+u=0$.  Therefore, it is very convenient
to use the subtracted propagator,
\eqn\subprop{\hat \C|P = \ln\hat \chi_{ij}(\nu^{(ij)}, \tau)= \C|P - C,}
which is modular invariant.  The expression \eprop\ can be written as a
Poincar{\'e}
  series,\foot{Recall that the Poincar{\'e} series associated
with a function $\psi$ defined over  $\C|F$  is
   $T\psi(\tau)=\sum_{\gamma\in
  \Gamma_\infty\backslash\Gamma} \psi(\Imagi\, \gamma z)$
   for $\tau=\tau_1+i\tau_2\in {\cal
  H}=\{\tau_2=\Imagi\, \tau > 0\}$
  and
  $\Gamma_\infty=
  \left\{\pm \pmatrix{1&n\cr 0&1\cr}, n\in\ZZ\right\}$.}
\eqn\ePoincareSeries{
\hat {\C|P}(\nu|\tau)=\sum_{p=1}^\infty {1\over p^2}\sum_{\gamma\in
  \Gamma_\infty\backslash\Gamma} \psi(\gamma(\nu),\gamma(\tau)),\quad \hbox{
  with } \psi(\nu,\tau) = {\tau_2\over 2\pi}\,  e^{-2i\pi p \hat\nu_2} }
where $\hat\nu_2 = \nu_2/\tau_2$ and the
$\Sl2$ transformation acts on $\nu$ and $\tau$ by
\eqn\sltwo{\tau \to {a\tau + b \over c\tau + d}, \qquad \nu \to
{\nu \over c\tau + d},}
   where $a,b,c$ and $d$ are integers and $ad-bc=1$.

 We will also need to express the propagator as a Fourier series
 in powers of $e^{2i\pi \tau_1}$, which has the form,
\eqn\eFourier{
\hat \C|P(\nu|\tau) ={\tau_2\over 4\pi}\sum_{n\neq 0} {1\over
n^2} e^{2i\pi n \hat\nu_2}+{1\over 4} \sum_{m\neq0\atop
k\in\ZZ}\, {1\over |m|} e^{2i\pi m(k\tau_1+ \nu_1)}\ e^{-2\pi \tau_2 |m|
 \left|k-\hat\nu_2\right|}
}

In analyzing the singular terms in the amplitude it will be
important to make use of the leading contribution to this expression for the
propagator at large values of $\tau_2$,
\eqn\asymdef{\hat \C|P^\infty(\nu|\tau) =
{\tau_2\over 4\pi}\sum_{n\neq 0} {1\over n^2} e^{2i\pi n
\hat\nu_2} ={\pi\tau_2\over 2}\left( \hat\nu_2^2  -  |\hat\nu_2| + {1\over 6}\right).}

\subsec{The diagrammatic rules}

The calculation of $I^{(m,n)}$ in \genderiv\ involves  integration of  powers
of the propagators, $\hat \C|P(\nu^{(ij)}|\tau)$, contracted between
various combinations of the points  $i,j$ ($i,j\in\{1,2,3,4\}$)
which are the locations of the vertex operators.

It is easy to deduce a set of diagramatic  rules  at any given
order.
 A term of order $\Delta^n$ (where each power of $\Delta$ may be any of the
 three $\Delta_r$'s with $r=s,t,u$)
contains a product of $n$ propagators which join pairs of points
(which we will call `vertices')
 $i,j$($=1,\dots,4$) with positions
$\nu^{(i)}$ that are to be integrated over the torus. We  will
represent each vertex of a diagram by a dot and each  propagator
linking  two vertices  by a line.
 The complete $n$th order
contribution  requires a sum over all ways  in which the
propagators can join the vertices.
For any  term in this sum every
 vertex that is not connected to any propagator contributes  a factor
of  $\int_{\C|T} d^2\nu_i/\tau_2=1$.

  More generally, we need to isolate divergent contributions by
dividing the $\tau$ integration domain  into two regions,
\eqn\totdomain{\C|F=  \C|F_\L + \C|R_\L.}
 The domain  $\C|F_\L$ defines the
`restricted' fundamental domain of the $\tau$
plane in which  $\tau_2\le \L$,
whereas the domain  $\C|R_\L$ defines a semi-infinite rectangle
in the $\tau$ plane, in which  $\tau_2\ge
\L$. As stated earlier, the
 terms $I_{nonan}^{(m,n)}$
  that have threshold singularities  at vanishing Mandelstam invariants
arise from the domain
$\C|R_\L$ and will be dealt with separately by integrating
over this large-$\tau_2$ region.

For the finite contributions that come from the domain
$\C|F_\L$  the integrations over the  positions, $\nu^{(i)}$
enforce overall conservation of the discrete momentum $p=m\tau+n$
in any diagram.  This means, for example, that any propagator
with a   free end-point gives a vanishing contribution since it
has been
 normalized to have a vanishing zero mode.
Therefore, non-zero contributions only come from diagrams in
which two or more propagators end on every vertex. Various
combinatorial factors are associated with each diagram and will
be described for each case separately.

\subsec{The threshold term, $I_{nonan}$}

The lack of analyticity of the low energy expansion of
 the one-loop amplitude \eLoop\  due to the logarithmic
 thresholds makes the integral representation \eLoop\
 ill-defined.  Since there is no region
 of the Mandelstam invariants in which the
amplitude is real the only way of making sense of the integral is
to  decompose the integration domain into three domains, $\calT_{st}$,
 $\calT_{tu}$ and $\calT_{us}$, so that the
amplitude is separated into real analytic
terms that have thresholds in the $(s,t)$, $(t,u)$ and
$(u,s)$ channels, respectively.
The integral representation for
 each of these terms can then be defined in the region of physical
 scattering, $s>0$; $t,u<0$,  by analytic
 continuation.  For example, the $(s,t)$ term is defined by
 continuation from the region $s,t<0$ (with $u=-s-t>0$) where it
 is real.  This decomposition follows very
naturally in an operator construction of the loop amplitude but
 does not  manifestly preserve modular invariance
 \refs{\rfDhokerPhongAnalytic}.

The leading logarithmic singularity in $I$ is the leading term in
$\lim_{s,t\to 0} \int (e^{\calD} -1)$.  This can be extracted by
first differentiating the integral representation with respect to $s$,
\eqn\leadt{\eqalign{\partial_s I_{nonan\, 1} = &\lim_{s,t\to 0}
 \int_\C|F {d^2\tau\over \tau_2^2}    \int_{\C|T}
\prod_{i=1}^3 {d^2\nu^{(i)}\over \tau_2} \,4\partial_s \calD
 \, e^{\calD}\cr
 = &\lim_{s,t\to 0} \int_\C|F {d^2\tau\over \tau_2^2}    \int_{\C|T}
\prod_{i=1}^3 {d^2\nu^{(i)}\over \tau_2} \, (4\Delta_s -4\Delta_u)
 \, e^{\calD}  .\cr}}
The contribution from the domain $\C|F_\L$ vanishes due to
the integration over the $\nu^{(ij)}$.  However, the region
$\C|R_\L$ leads to a nonzero result.  In this region we can
approximate $\calD$ by using the asymptotic expression for the
propagator \asymdef\
which is proportional to $\tau_2$.  In the term with thresholds in the
$(s,t)$ channels  the
variables $\nu_2^{(i)}$ are ordered in such a manner that the
rescaled   variables,
\eqn\rescnu{\omega_i = {\nu^{(i)}_2\over \tau_2},}
span the range
\eqn\doner{\calT_{st}: \quad 0\le \omega_1 \le \omega_2\le \omega_3\le
\omega_4 =1}
 (where we have used the conformal symmetry to fix $\nu^{(4)} = \tau$).
The various permutations of
this ordering are relevant in the $(t,u)$ and $(u,s)$ regions so
that the whole range, $0\le \omega_i\le 1$ is covered by adding
the three regions, $\calT_{st}$, $\calT_{tu}$ and $\calT_{us}$,  together.
In terms of these variables we have, in the  region $\calT_{st}$,
\eqn\delreg{\eqalign{\calD= \calD(s,t)=&\lim_{\tau_2\to\infty}
 \alpha's(\Delta_s - \Delta_u) +
 \alpha' t(\Delta_t-\Delta_u)\cr
 =& \alpha'\pi\tau_2
\left(s \omega_1 (\omega_3 -\omega_2) + t(\omega_2-\omega_1) (1
-\omega_3)\right)\cr
=& \alpha'\pi\tau_2 {\cal Q}(s,t),\cr}}
where
\eqn\qdef{{\cal Q}(s,t)
= s \omega_1 (\omega_3 -\omega_2) + t(\omega_2-\omega_1) (1
-\omega_3).}
Similar expressions  for the functions $\calD(t,u)=
t(\Delta_t - \Delta_s) + u(\Delta_u-\Delta_s)$ and $\calD(u,s)=
u(\Delta_u - \Delta_t) +s(\Delta_s-\Delta_t)$,
define  $\calD$ when expressed in the $\calT_{tu}$ and $\calT_{us}$ regions.
In the $\C|R_\L$ domain of \leadt\ the $\tau_1$ integration
is trivial since the integrand has no $\tau_1$ dependence.  The
$\tau_2$ integration (from $\L$ to $\infty$) simply gives
\eqn\rlamint{\eqalign{\partial_s I_{nonan\, 1} =  &4\pi  \int_\L^\infty
{d\tau_2\over \tau_2}
\prod_{0\le\omega_1\le\omega_2\le\omega_3\le1}
 d\omega_i  \,  \omega_1 (\omega_3 -\omega_2)\,
  e^{\alpha'\pi\tau_2 {\cal Q}(s,t)}\cr
 =&4\pi\int \prod_{0\le\omega_1\le\omega_2\le\omega_3\le1}
 d\omega_i  \,  \omega_1 (\omega_3 -\omega_2)\,
    \left(-\gamma-\ln (-\alpha'\pi{\cal Q}(s,t))-\ln\L\right)
  + o(s).\cr}}
The $\ln \L$ terms cancel out in the complete contribution to
$I^{(1,0)}= \partial_sI - \partial_u I$.
It it is easy to integrate \rlamint\
 together with  the corresponding expression
for $\partial_t I_{nonan\, 1}$, giving
\eqn\qresult{\eqalign{{1\over 4\pi}
I_{nonan\, 1} = &\int \prod_{0\le\omega_1\le\omega_2\le\omega_3\le1}
 d\omega_i  \, {\cal Q}(s,t)\,  \ln {\cal Q}(s,t)
  + \int \prod_{0\le\omega_3\le\omega_2\le\omega_1\le1}
 d\omega_i  \, {\cal Q}(t,u)\,  \ln {\cal Q}(t,u) \cr
 & +
 \int \prod_{0\le\omega_2\le\omega_1\le\omega_3\le1}
 d\omega_i  \, {\cal Q}(u,s)\,  \ln {\cal Q}(u,s).\cr}}
The scale of the logarithm cancels out of the sum of terms in the
full expression.   This threshold term is exactly the same   as that obtained from
 the one-loop calculation of the four-graviton amplitude in either of the
  type II supergravity theories
  in ten dimensions \refs{\rfGreenCargese,\rfRussoTseytlin}.
The corresponding discussion of the higher-order threshold term,
$I_{nonan\, 2}$, which is intrinsically stringy since it involves higher powers
of $\alpha'$,   will be given in the appendix.

Such threshold terms are contained in the large-$\tau_2$ region of the
integration over moduli space, which means that they are contained  in
the coefficients $I^{(m,n)}_{\C|R_\L}$ that are
defined by integration over the
domain $\C|R_\L$. So long as $1<m+n<4$
it will be sufficient to substitute the
asymptotic form of the propagator which will produce
contributions of the form,
\eqn\alphf{
\eqalign{
\lim_{s,t\to 0} &
I_{\C|R_\L}^{(m,n)}(s,t) = \lim_{s,t\to 0} \int_{\C|R_\L}
{d^2\tau\over \tau_2^2} \int_{\C|T} \prod_{i=1}^3 {d^2\nu^{(i)}\over \tau_2}
\, (4\Delta_s-4\Delta_u)^m(4\Delta_t-4\Delta_u)^n  \, e^{\calD} \cr
= &\lim_{s,t\to 0} \int_\L^\infty d\tau_2 \tau_2^{m+n-2} \int_\C|T
\prod_{i=1}^3 d\omega_i
\,\left(4\pi\partial_s{\cal Q}\right)^m \left(4\pi\partial_t{\cal Q}\right)^n
\, e^{\alpha'\pi\tau_2 {\cal Q}}  \cr
= & \lim_{s,t\to 0} \left(\int_0^\infty d \tau_2 -\int_0^\L d\tau_2\right)
 \tau_2^{m+n-2} \int_\C|T \prod_{i=1}^3
 d\omega_i   \,\left(4\pi\partial_s{\cal Q}\right)^m \left(4\pi\partial_t{\cal
Q}\right)^n \, e^{\alpha'\pi\tau_2 {\cal Q}} \cr
= &\lim_{s,t\to 0} I_{nonan}^{(m,n)}(s,t) - 2\times (4\pi)^{m+n} {\L^{m+n-1}
\over m+n-1}\cr
& \partial_s^m\partial_t^n \left.\sum_{p+q=m+n}  {p! q! \over
(2m+2n+3)!} \left[s^p t^q + (-1)^q s^p (s+t)^q + (-1)^p (s+t)^p t^q \right]\right|_{(s,t)=(0,0)}
, \cr
}}
with $p,q\ge 0$.  The $\L$-dependent term in this expression
will cancel with a
term that arises from the integration over the domain
$\calF_\L$.  When $m+n=4$ it is no longer adequate to use
the leading $\tau_2$ contribution to the propagators and a
sub-leading contribution of the form $\ln \L$ arises.  This
leads to the term $I_{nonan\, 2}$, as will be seen in more
detail in section 4.3 and the appendix.

\newsec{The analytic terms, $I_{an}^{(m,n)}$}

The analytic terms are extracted from the integration over $\calF_\L$
which is  finite.
In this domain we can first perform the $\nu^{(i)}$ integrals to
obtain a density on the moduli space  and then
integrate this  over $\tau$ and $\bar \tau$.

The first term in the expansion of~\eF\ using \genderiv\ is
 the trivial constant term.  The result of the integrations is
 simply the finite volume of $\C|F$.  This defines the first
constant in \loopexp,
\eqn\const{I^{(0,0)} =
a = \int_\C|F {d^2\tau \over \tau_2^2} ={\pi \over 3} ,}
which is the well-known coefficient of the loop contribution to the
$\calR^4$ term.

The next terms  in the expansion  are $I^{(1,0)}$ and
$I^{(0,1)}$ which are given by \leadt.
 As remarked earlier, the $\nu^{(i)}$ integrations in \leadt\  cause the
integrand to vanish in the domain $\calF_\L$
 and  the integral  only contributes to $I^{(1,0)}_{nonan}$.

\subsec{Terms of order $s^2$}

\ifig\fone{ The diagram that contributes to $I_{an}^{(m,n)}$ with $m+n=2$.}
{\epsfbox{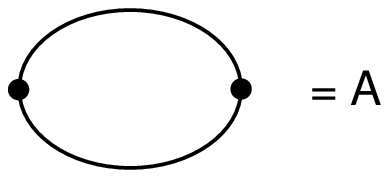}}

The only  non-vanishing contribution to the integrand of $I_{an}$
 at order ${\alpha'}^2$ is the bubble diagram of \fone.  This term
 multiplies  $(s^2+t^2+u^2)$ in the expansion \anloop\ and
  therefore contributes to
  $I_{an}^{(2,0)}$, $I_{an}^{(1,1)}$ and $I_{an}^{(0,2)}$.
  The density on moduli space  arising from \fone\ is
\eqn\twobub{
\eqalign{
A(\tau,\bar\tau)
= & \int_{\C|T}
 {d^2\nu^{(i)} d^2\nu^{(j)}\over \tau_2^2} \
 \left[\ln\hat\chi(\nu^{(ij)}|\tau)\right]^2\cr
 =& {1\over 16\pi^2} \sum_{(m,n)\neq (0,0)} {\tau_2^2\over |m\tau+n|^2} =
{1\over 16\pi^2} Z_2(\tau,\bar\tau).\cr
}}
The function $Z_2$ an Epstein zeta function which  is an example of a non-holomorphic
 Eisenstein series\foot{This is related to  the function $E_s$ in
 \refs{\rfTerrasBook} by $Z_s = 2\zeta(2s)\, E_s$.}.
 More generally these are non-holomorphic modular
functions that are also eigenfunctions of the Laplace operator in
the fundamental domain of $\Sl2$,
\eqn\eigendef{\nabla^2 Z_s \equiv 4\tau_2^2{\partial^2\over \partial \tau
\partial\bar\tau} Z_s = s(s-1) Z_s.}
These functions have $\tau_2$ expansions in which there are two
power-behaved terms together with an infinite set of
exponentially suppressed, non-perturbative, terms,
\eqn\eEsexp{Z_s= \sum_{(m,n)\ne (0,0)}{\tau_2^s\over |m\tau + n|^{2s}}
 = 2\zeta(2s) \tau_2^s +
2\pi^{1/2}{\Gamma(s-1/2)\over \Gamma(s)}\zeta(2s-1)\tau_2^{1-s} +
O(e^{-2\pi \tau_2}).}

Diagrams of this type with vertices at $(1,2)$,
$(3,4)$, $(1,3)$ and $(2,4)$  contribute equally  to
$I^{(2,0)}_{an}$. Integrating
  \twobub\  over the restricted fundamental domain,
  $\C|F_\L$, and summing over these four contributions
  gives
\eqn\essquare{
I_{\C|F_\L}^{(2,0)} =4\times {1\over \pi^2}
\int_{{\cal F}_\L} {d^2\tau\over \tau_2^2}
Z_2(\tau,\bar\tau) ,}
where the   factor of 4 comes from
$\partial_s^2(s^2+t^2+(s+t)^2)$.
This expression is easily integrated by substituting
$Z_2=\nabla^2 Z_2/2$ using
\eigendef\ so that the integrand is a total derivative and
\essquare\
reduces to an integral over the boundary.  The restricted fundamental
domain has a single boundary which is at $\tau_2 =\L$ and
the result is
\eqn\gausslas{
I_{\C|F_\L}^{(2,0)} ={2^6\;\pi^2\over 6!}\L +
  O(\L^{-2}).}
This $\L$-dependent term cancels the corresponding $\L$-dependence  arising
from $I^{(2,0)}_{\calR_\L}$  which is given by \alphf.
Since there is no residual $\L$-independent piece
we conclude  that   $I^{(2,0)}_{an}=0$, which implies that
\eqn\cdef{b=0.}
This  means that there is no $s^2$ term in the expansion
\loopexp.

In a similar manner it is easy to verify  that the cross term
$I_{an}^{(1,1)}$ also vanishes, which is consistent with \sderivs.

\subsec{Terms of order $s^3$}

\ifig\ftwo{Diagrams that contribute to $I_{an}^{(m,n)}$ with $m+n=3$.}
{\epsfbox{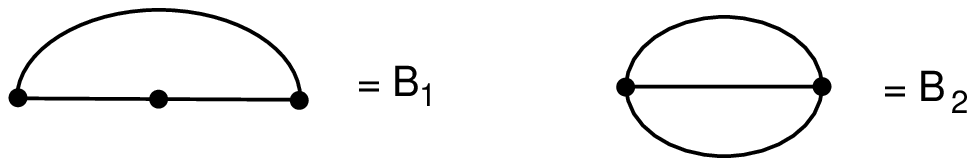}}

The diagrams in \ftwo\ are the ones that survive the $\nu^{(ij)}$
integrations.  These
contribute to terms in the expansion \anloop\  with coefficient
 ${\alpha'}^3(s^3+t^3+u^3)$.
The first diagram is the product of three propagators joining
 three distinct vertices and
gives contributions to the
integrand of $I_{\C|F_\L}^{(2,1)}$ (and $I_{\C|F_\L}^{(1,2)}$)
of the form ($k>j>i$)
\eqn\escubeb{ B_1(\tau,\bar\tau)
 ={1\over(4\pi)^3} \sum_{m,n} {\tau_2^3\over |m\tau+n|^6} =
{1\over (4\pi)^3}Z_3(\tau,\bar\tau) . }
 This is again  a
non-holomorphic Eisenstein series
 satisfying \eigendef, here  with $s=3$,  so it is an eigenfunction
 of the Laplace equation with   $s(s-1)=6$.
 The integration of the density \escubeb\ over the the restricted fundamental
 domain   can again be performed using Gauss' law.  This  gives,
\eqn\ittwoone{
I_{\C|F_\L}^{(2,1)}(B_1) =-{2^{10}\;\pi^3\over 3\cdot 8!} \L^2 + O(\L^{-3}),}
where the factor of $-6$ arises from
$\partial_s^2\partial_t(s^3+t^3-(s+t)^3)$.

The second  diagram of \ftwo\ involves only two distinct vertices
and potentially gives a contribution to the integrand of both
$I_{\C|F_\L}^{(3,0)}$ and  $I_{\C|F_\L}^{(2,1)}$
 (as well as $I_{\C|F_\L}^{(0,3)}$ and  $I_{\C|F_\L}^{(1,2)}$)
  of the form
\eqn\escubetb{\eqalign{
B_2(\tau,\bar\tau) = & \int
{d^2\nu^{(i)}d^2\nu^{(j)}\over \tau_2^2}
\left(\hat\C|P(\nu^{(ij}|\tau)\right)^3\cr
 = &{1\over
  (4\pi)^3} \sum_{(m,n),(k,l),(p,q)\neq(0,0)} \ \tau_2^3\
{\delta(m+k+p)\delta(n+l+q)\over |m\tau+n|^2 |k\tau+l|^2
|p\tau+q|^2} .\cr}}
In fact, $I^{(3,0)}$ involves the combination $\partial_s^3(s^3+t^3 - (s+t)^3) = 0$,
so it automatically vanishes as in \loopexp.
However, the integrand of $I^{(2,1)}$ is proportional to $B_2$.

Unlike the earlier examples, this expression is not  an eigenfunction of the
Laplacian on the fundamental domain so a new idea is needed in
 order to perform the integration over $\C|F_\L$.  We will
 make use of the well-known `unfolding procedure' by
 using the representation of the propagator by a
  Poincar{\'e} series \ePoincareSeries.
This relates the integral of $\psi\times f$ over
$\C|F$ (where $\psi$ is  any Poincar{\'e} series) to an integral
over the semi-infinite line,
\eqn\tricky{\int_{\cal F}
{d^2\tau\over \tau_2^2} \psi(\tau)f(\tau) = \int_{t=0}^\infty
{dt\over t^2} \psi(t) (C f)(t),}
where the expression $(Cf)$ is the  zero $\tau_1$ mode of the function
 $f(\tau)$,
\eqn\unfold{Cf(\tau_2) = \int_{-1/2}^{1/2} d\tau_1 f(\tau).}
The relationship  \tricky\ is derived by   making use of the identities,
\eqn\useuden{\int_{\Gamma\backslash {\cal H}}
\sum_{\Gamma_\infty\backslash\Gamma} =
\int_{\Gamma_\infty\backslash{\cal H}}=\int_{\tau_2>0}.}
Since the integration in  \escubetb\ is over the restricted
fundamental domain, $\calF_\L$,  some
care has to be taken in using the unfolding
procedure.  For an integral such as \escubetb, which diverges like a power
of $\L$ as $\L\to \infty$, it turns out to be  consistent to
 simply set $f(\tau)=0$ for $\tau_2\ge \L$,  which cuts off
the divergence at $\tau_2\to \infty$.\foot{ Although this cut-off leads to the
  correct answer when the integral grows as a power of $\L$, more care is
  needed in regularizing logarithmic growth of the kind we will meet in
  section 4.3. In that case the integral diverges at the points on the
  $\tau_2=0$ axis that are the images  under $SL(2,\ZZ)$ transformations of
  the point $\tau_2\to\infty$.}
Using this procedure we can express the contribution of  \escubetb\ to
$I_{\C|F_\L}^{(2,1)}$  in the form,
\eqn\escubetrick{
\eqalign{
I_{\C|F_\L}^{(2,1)} (B_2) &
= -6\times {64\over 3}\int_{\C|F_\L} {d^2 \tau\over
  \tau_2^2} \, B_2(\tau,\bar\tau)  \cr
&=-128 \int_\C|F {d^2\tau\over \tau_2^2} \int
{d^2\nu^{(1)}d^2\nu^{(2)}\over \tau_2^2} \left(\hat
  \C|P(\nu^{(12)}|\tau)\right)^3 \cr
&= -128\sum_{p=1}^\infty {1\over p^2} \int_0^\L {dt\over t^2} \int
{d^2\nu^{(1)} d^2\nu^{(2)}\over t^2}\; {t\over 2\pi} e^{-2i\pi p \hat\nu_2} \;
C(\hat{\cal P}^2) ,\cr }}
where the overall factor of $-6$ comes from
  $\partial^2_s\partial_t (s^3+t^3-(s+t)^3)$.
The expression for the  integral
over $\nu^{(1)}_1$ and $\nu^{(2)}_1$ of the  zero $\tau_1$  Fourier mode is
\eqn\ecdefg{\int_{-\half}^{\half} d\nu^{(1)}_1 d\nu^{(2)}_1
C(\calP^2) = {t^2\over 16\pi^2} \left(\sum_{n\ne 0} {1\over n^2}
e^{2\pi i n\hat \nu_2^{(12)}}\right)^2 + {1\over 16} \sum_{m\ne
0\atop k} {1\over m^2} e^{-4\pi t|m||k-\hat \nu_2^{(12)}|}.}
Substituting the first term on the right-hand side into \escubetrick\ leads to the
$\L$-dependent term,
\eqn\lambdadep{
\eqalign{
I^{(2,1)}_{1\,\C|F_\L}(B_2) &=-128\times {\L^2\over
    2}\; \int_{-\half}^\half d\hat\nu_2^{(1)} d\hat\nu_2^{(2)} \left( {1\over
      4\pi} \sum_{n\neq 0} {1\over n^2} e^{2i\pi n\hat
      \nu_2^{(12)}}\right)^3\cr
& =-{\L^2\over \pi^3}
  \sum_{n_i\in Z\backslash\{0\}} {\delta(n_1+n_2+n_3)\over n_1^2
    n_2^2 n_3^2} =-{2^8\; \pi^3\over 3\cdot 8!}  \L^2.}}
Substitution of the second term on the right-hand side of
\ecdefg\ into \escubetrick\ gives the $\L$-independent term,
\eqn\lambinde{\eqalign{
I^{(2,1)}_{2\, \C|F_\L}(B_2)  &=-{4\over \pi}\sum_{p=1}^\infty {1\over
  p^2}\int_0^\infty {dt\over t} \int_0^1 d\hat\nu^{(1)}_2  d\hat\nu^{(2)}_2
\sum_{m\neq 0\atop k} {1\over m^2} e^{-2i\pi p \hat\nu_2} e^{-4\pi t|m| |k
  -\hat\nu_2|}\cr
&=-{4\over \pi^2}\sum_{p=1}^\infty\sum_{m\neq 0} {1\over p^2} {1\over
m^2}\int_0^\infty {dt\over p^2+t^2}=-6\times {2\over 3\pi} \zeta(2)\zeta(3) .\cr}}

The sum of the $\L$-dependent terms, $I^{(2,1)}_{1\,\calF_\L}(B_2)
+I^{(2,1)}_{\calF_\L}(B_1)$  (see \lambdadep\ and \ittwoone),
again cancels with the corresponding
  term in the integration over the domain $\C|R_\L$
  ($I_{\C|R_\L}^{(2,1)}$, in~\alphf).
   However, in this case
  there is also a finite contribution to $I^{(2,1)}_{2\, \C|F_\L}$, which
  determines the coefficient of the ${\alpha'}^3(s^3 + t^3 + u^3)$ term in
   \loopexp\ to be
\eqn\ddef{
c ={2\over 3\pi} \zeta(2)\zeta(3).
}

\subsec{Terms of order $s^4$}

\ifig\fthree{ The set of  diagrams that
 contribute to $I^{(m,n)}$ with $m+n=4$.}
{\epsfbox{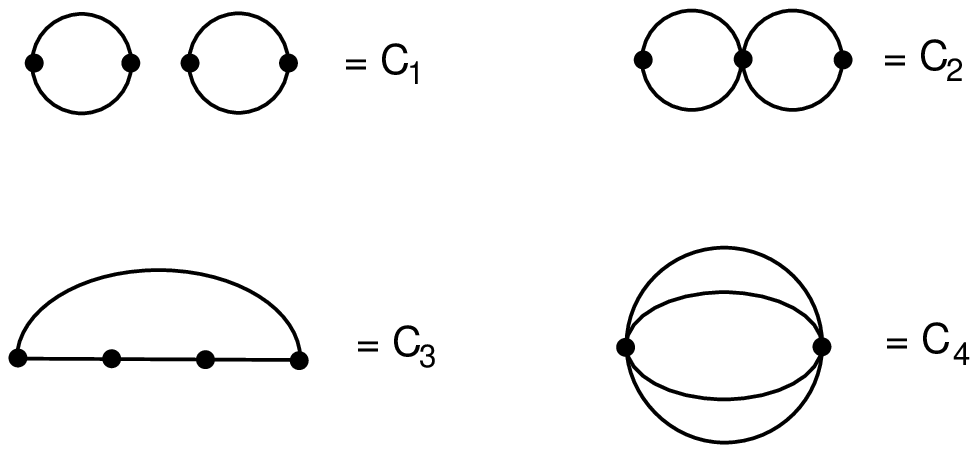}}
The  four kinds of
diagrams that give non-zero contributions proportional to  ${\alpha'}^4\,
 (s^4+t^4+u^4)$ in the expansion \anloop\ are shown in \fthree.
 These each have four propagators and
 contribute to $I_{an}^{(m,n)}$ with $m+n=4$.
Upon evaluating the $\nu^{(r)}$ integrations these give the
following  densities for  the moduli space integrals
\eqn\figcone{
C_1(\tau,\bar\tau) =
{1\over (4\pi)^4}(Z_2)^2 ,}
\eqn\figctwi{C_2(\tau,\bar\tau)  =
{1\over (4\pi)^4}(Z_2)^2  ,}
\eqn\figcthree{
C_3(\tau,\bar\tau) =
{1\over (4\pi)^4}Z_4,}
\eqn\figcfour{
\eqalign{
C_4(\tau,\bar\tau) &=  \int {d^2\nu^{(i)}d^2\nu^{(j)}\over \tau_2^2}
\left(\hat\C|P(\nu^{(ij)}|\tau)\right)^4 \qquad (j>i)\cr
&= {1\over (4\pi)^4}\sum_{(m,n)\neq(0,0)\atop(p,q)\neq(0,0)}\!\!
\sum_{(r,s)\neq(0,0)\atop(v,w)\neq(0,0)}\tau_2^4\
{\delta(m+p+r+v)\delta(n+q+s+w)\over |m\tau+n|^2 |p\tau+q|^2 |r\tau+s|^2
  |v\tau+w|^2}\ .
}}

The term $C_3(\tau,\bar\tau)$  is once again a non-holomorphic
Eisenstein series
which can be integrated over the restricted fundamental domain using,
\eqn\eisenfour{\int_{{\cal F}_\L}
{d^2\tau\over \tau_2^2} Z_4 ={2\over 3} \zeta(8) \L^3 + O(\L^{-4}).}
Inserting the appropriate combinatoric factors
 gives  rise to the $\L$-dependent contributions,
\eqn\lambdep{I^{(3,1)}_{an}(C_3) =0 ,\quad
  I^{(2,2)}_{an}(C_3)={2^{10}\cdot 3\;\pi^4\over 10!}\L^3,\quad
  I^{(4,0)}_{an}(C_3)={2^{10}\cdot 3\;\pi^4\over 10!}\L^3.}

The integration of the expressions $C_2$ and $C_3$ over the restricted fundamental
domain  involves the
integration of the square of an Eisenstein series, $(Z_2)^2$.  This
can be evaluated by using Green's theorem in the fundamental
domain.  For general real values of $s,s'>1/2$ this states that
\eqn\terrasthm{
{1\over 4\zeta(2s)\zeta(2s')}\int_{{\cal F}_\L} {d^2\tau\over \tau_2^2}
Z_s Z_{s'} =
{\L^{s+s'-1}\over s+s'-1} - {\L^{1-s-s'}\over s+s'-1}
\phi(s)\phi(s')+ {\L^{s-s'}\over s-s'} \phi(s') -
{\L^{s'-s}\over s-s'} \phi(s) +o(1),
}
where
\eqn\phidef{
\phi(s) = {\zeta(2s-1)\Gamma(s-1/2)\over
\pi^{s-1/2}}{\pi^s\over\zeta(2s)\Gamma(s)}.
}
The symbol $o(1)$ means that the remainder goes to zero when
$\L$ becomes infinite (for a more general statement see
exercise 12, page 216,  of volume I of  \refs{\rfTerrasBook}).
It follows from this that
\eqn\eLogDiv{
{1\over 4\zeta(4)^2}\int_{{\cal F}_\L} {d^2\tau\over \tau_2^2} (Z_2)^2= {1\over
3}\L^3 + \pi {\zeta(3)\over \zeta(4)}\ln\,\L
- \phi'(2).}
 The last term gives a $\L$-independent
contribution to the coefficient $d$ in~\loopexp.  The other terms
in \eLogDiv\ give another  contribution that behaves as $\L^3$ as well as
 a new $\L$-dependent term proportional to
$\ln\, \L$. Such a term is implied by the presence of a new
logarithmic threshold  of order ${\alpha'}^4 \,s^4\, \ln s$
which is contained in $I_{nonan\, 2}$ that is evaluated in the appendix.
 Taking into account the combinatorial factors, the
contribution of these $\L$-dependent terms is
\eqn\clams{\eqalign{
I^{(3,1)}_{an}(C_1)&=I^{(3,1)}_{an}(C_2)=0\cr
I_{an}^{(2,2)}(C_2)&=2I^{(2,2)}_{an}(C_1)={2^9\cdot 7\;\pi^4\over 10!}
\L^3 + {48\over\pi^3} \zeta(3)\zeta(4) \ln\, \L, \cr
I_{an}^{(4,0)}(C_2) &=2I^{(4,0)}_{an}(C_1)={2^{10}\cdot 7\;\pi^4\over
  10!}\L^3 +{96\over\pi^3}\zeta(3)\zeta(4)  \ln\, \L   .}}

The last remaining term to consider is $C_4$.  As with the
two-loop term~\escubetb\ this gives an expression which is not
an eigenfunction of the Laplacian on the fundamental domain. Once again
the $d^2 \tau$  integration may be performed by using the unfolding
procedure as in the previous sub-section. However, in
this case we have to take greater care of the divergence of the
integrand at $\tau_2 =\L\to \infty$ (as pointed out in the  footnote in section 4.2).
  The integral can be rendered finite by subtracting
a suitable linear combination of $Z_4$ and $Z_2^2$ from the
integrand. Consider, for example,
\eqn\cfourty{I^{(4,0)}_{\calF_\L}(C_4) =
 48\times {64\over 3} \int_{\C|F_\L} {d^2\tau\over \tau_2^2} \int
{d^2\nu^{(1)}d^2\nu^{(2)}\over \tau_2^2} \
\left(\hat\C|P(\nu^{(12)}|\tau)\right)^4 ,}
where   the factor of $48$ comes from $\partial_s^4(s^4+t^4+(s+t)^4)$.
It is easy to extract the terms in the integrand that are divergent in
the limit
$\L\to \infty$ from explicit form of
$\hat\C|P(\nu^{(12)}|\tau)$ given  in \eFourier\ and \asymdef.  This
gives explicit $\L^3$ and $\ln\L$ terms that can be
subtracted in a modular invariant manner
 by defining a regularized value of $I^{(4,0)}(C_4)$,
\eqn\unfort{I^{(4,0)}_{\rm Reg}(C_4)
=I^{(4,0)}_{\calF_\L}(C_4) - I^{(4,0)}_{div}(C_4),}
where
\eqn\ionedef{I^{(4,0)}_{div}(C_4) =
1024
  \int_{\C|F_\L} {d^2\tau\over \tau_2^2} \int
  {d^2\nu^{(1)}d^2\nu^{(2)}\over \tau_2^2} \
  \left(- {2\over (4\pi)^4} Z_4 +
  {3\over (4\pi)^4} (Z_2)^2\right),}
and the integrals of $Z_4$ and $Z_2^2$ are given in \eisenfour\
and \eLogDiv, respectively.

Since the expression \unfort\  is finite and its integrand is modular
invariant it is straightforward to evaluate  using the
unfolding procedure. This gives
\eqn\esFourTrick{\eqalign{
I^{(4,0)}_{\rm Reg}(C_4)= &
1024\sum_{p=1}^\infty {1\over p^2}\; \int_0^\L
{dt\over t^2}\;\int {d^2\nu^{(1)}d^2\nu^{(2)}\over t^2} \
{t\over 2\pi} e^{-2i\pi p\hat\nu_2^{(12)}} \;
\left[C\left(\hat\C|P(\nu^{(12)})^3\right)
\right.  \cr
& \left.
+ 2 \int {d^2\nu^{(3}) d^2\nu^{(4)}\over t^2}
  C\left(\hat\C|P(\nu^{(23)})\, \hat\C|P(\nu^{(34)})\,
\hat\C|P(\nu^{(41)})\right) \right. \cr & \left.
  -  {3\over (4\pi)^2}\int {d^2 \nu^{(3)} d^2\nu^{(4)} \over t^2}
   C\left(\hat\C|P(\nu^{(12)})Z_2\right)\right]
  .\cr}}
This term can be evaluated by using the explicit definitions of $\hat {\cal P}$ and
$Z_2$, giving the $\L$-independent result,
\eqn\eSfourFinite{
\eqalign{
I^{(4,0)}_{\rm Reg}(C_4) =&-{24\over \pi^3}\zeta(3) \sum_{p\neq0, n\neq0\atop p\neq n} {\ln|p-n|\over p^2n^2} \cr
& + {8\over \pi} \sum_{m_i\neq 0,\; m_1+m_2+m_3=0\atop k_i\in \ZZ,\;
m_1k_1+m_2k_2+m_3k_3=0} {1\over |m_1m_2m_3|}\sum_{p=1}^\infty {1\over p^2}
\int_0^\infty {dt\over t}\cr
&   \int_0^1 d\hat\nu_2^{(1)}d\hat\nu_2^{(2)} \exp\left(-2i\pi p
\hat\nu_2^{(12)}-2\pi t \sum_{i=1}^3 |m_i| |k_i-\hat\nu_2^{(12)}|\right) \
.\cr
}}
The $\L$-dependent terms are contained in
\eqn\esFourLambda{
I^{(4,0)}_{div}(C_4)=2I^{(2,2)}_{an}(C_4)= {2^9\cdot 5\;\pi^4\over 10!}
  \L^3 + {48\over \pi^3}\zeta(3)\zeta(4)\ln\L -{48\over \pi^4} \zeta(4)^2 \phi'(2)\ .
}
The $\L^3$ term is
connected, as expected, to the presence of $I^{(4,0)}_{nonan\, 1}$ while the $\ln
\L$ term  is again connected to the appearance of $I_{nonan\,
2}^{(4,0)}$.

 The sum of the $\L^3$ contributions arising in
$I_{an}^{(4,0)}(C_1)$ and $I_{div}^{(4,0)}(C_4)$ indeed cancels the
contributions from the integration of $I^{(4,0)}_{nonan\, 1}$ over the
$\calR_\L$ domain in \alphf.  Similarly, the total coefficient of
$\ln \L$ arising in the sum of $I_{an}^{(2,2)}(C_1)$,
$I_{an}^{(2,2)}(C_2)$ and $I_{an}^{(2,2)}(C_4)$ is
\eqn\sumlog{
I^{(2,2)}_{an\;\ln}= {96\over \pi^3} \zeta(3)\zeta(4)\ln\L , }
which will be cancelled by the presence of the new threshold term
$I^{(2,2)}_{nonan\, 2}$. The  general expression for $I_{nonan\, 2}$ is fairly
complicated but we see from the appendix  that at $t=0$ it reduces   to
\eqn\inonann{I_{nonan\, 2}(s,t=0) = {4\over \pi^3}
\zeta(3)\zeta(4) (\alpha's)^4 \left(\ln\left( {\alpha' s\over \L} \right)+
\ln\left(-  {\alpha' s\over \L} \right)\right).}
 Taking four $s$ derivatives leads to the same
coefficient of $\ln \L$ as that in  \sumlog.

The finite term $I^{(4,0)}_{\rm Reg}(C_4)$ \eSfourFinite,
 together with the  $\L$-independent parts of
 \esFourLambda\ and $I^{(4,0)}(C_2)$ and $I^{(4,0)}(C_3)$
(which come from the finite last term  of \eLogDiv),
  determine  the value of the coefficient $d$ in the expansion of
the loop amplitude in the form,
\eqn\dres{\eqalign{
d=& - {4\over \pi^4} \zeta(4)^2 \phi'(2) - {1\over 2\pi^3} \zeta(3) \sum
{\ln|p-n|\over p^2 n^2} \cr
& + {1\over 6\pi} \sum_{m_i\neq 0,\; m_1+m_2+m_3=0\atop k_i\in \ZZ,\;
m_1k_1+m_2k_2+m_3k_3=0} {1\over |m_1m_2m_3|}\sum_{p=1}^\infty {1\over p^2}
\int_0^\infty {dt\over t}\cr
&   \int_0^1 d\hat\nu_2^{(1)}d\hat\nu_2^{(2)} \exp\left(-2i\pi p
\hat\nu_2^{(12)}-2\pi t \sum_{i=1}^3 |m_i| |k_i-\hat\nu_2^{(12)}|\right) \
.\cr
}}
We have not extracted the numerical value of this complicated
looking expression.

\newsec{Summary and conclusion}

In summary, we have determined the first few coefficients in the
expansion \loopexp\ of the four-graviton  one-loop amplitude in either of
the ten-dimensional type II string theories.  After explicitly
subtracting the non-analytic threshold terms $I_{nonan\, 1}$ and $I_{nonan\,
2}$,  we found that
\eqn\summres{a={\pi\over 3}, \qquad b=0, \qquad c={2\over 3\pi}
\zeta(2)\zeta(3),}
and $d$ is given by the expression \dres\ that we have not
evaluated.

These coefficients give a little more insight into the
structure of the low energy expansion of four-graviton interactions in the
 M theory effective action.  The leading term of this type is the
 $\calR^4$ term about which a great deal is known
\refs{\rfGreenGutperle,\rfGreenVanhoveMtheory,\rfGreenSethi,\rfBoris,\rfHerveAlex,\rfGreenGutperleVanhove}.
For example,
 in the ten-dimensional limit corresponding to the type IIB
 string theory,  it has dependence on the complex coupling,
  $\Omega = C^{(0)} + i e^{-\phi^B}$ (where $C^{(0)}$ is the \RR\ scalar and $\phi^B$
is the type IIB dilaton),  that enters by an overall factor of
$E_{3/2}(\Omega,\bar\Omega)$, where
$E_s$ is the modular invariant  Eisenstein series that is proportional to $Z_s$
 (see the footnote in section 4.1).
This  function has
an expansion for large $\Omega_2$ (weak coupling) that begins with
the tree-level term with coefficient $\zeta(3)$ in \eTreeExp\ and
is followed by a  one-loop term with a coefficient that is
precisely the value of $a$ in \summres.
There are no further perturbative terms in the expansion but there
is a precisely defined sequence of D-instanton contributions.

One method by which the exact form of the
 the $\calR^4$ interaction was determined
 \refs{\rfGreenGutperleVanhove} by calculating  the  one-loop contribution to
four-graviton scattering in eleven-dimensional supergravity
compactified on a two-torus.
Recently this method has been generalized to evaluate the two-loop
contribution in eleven-dimensional supergravity
which contributes at leading order in the low energy
expansion to the  $D^4 \, \calR^4$
  interaction, where the notation symbolically indicates four derivatives acting on four
  powers of the curvature.   In the limit that gives the ten-dimensional
  type IIB theory   the interaction is given by a term in the effective
  action density of the form \refs{\rfGreenKwonVanhove}
\eqn\exactin{\zeta(5)\, \C|V^{-5/2}\, E_{5\over
 2}\Omega,\bar\Omega) \, (s^2+t^2+u^2) \,
\calR^4}
(where the factors of $s^2$, $t^2$ and $u^2$ represent appropriate
derivatives acting on the curvature tensors).
 In this case the modular function $E_{5/ 2}$ has
an expansion for large $\Omega_2$ (weak coupling) that begins with
the tree-level term with coefficient $\zeta(5)$ in \eTreeExp\ and
is followed by a {\it two-loop term}  --- the one-loop
contribution is absent.
 Again there are no further perturbative string theory
contributions but there is an infinite series of D-instanton
contributions.  The vanishing of the one-loop contribution in \exactin\ is
 confirmed   by
our statement that the coefficient $b$ in \summres\ vanishes.

The value of $c$ in \summres\ is the coefficient of the one-loop contribution
to the  $(s^3+t^3+u^3) \, \calR^4$ interaction.  This is not a
term which has yet been motivated from any argument based on duality or supersymmetry.  In
particular, it is not yet clear how this
term packages with the tree-level $\zeta(3)^2$
term in \eTreeExp\ to make a modular invariant expression in the
type IIB limit.

More generally, one might ask whether there is a simple modular
invariant expression for the complete four-graviton amplitude that
generalizes the tree amplitude \eTreeta.  An
obvious candidate is   obtained by replacing the
coefficients $2\zeta(2n+1)$ in the tree amplitude
\eTreeta\ by $\tau_2^{-2n-1}\, 2\zeta(2n+1)E_{n+{1/ 2}}$
\refs{\rfRusso,\rfGreenMoore}. The resulting amplitude   has $s$, $t$ and $u$-channel
poles at values corresponding to the mass of every excited state of all the
$(p,q)$ D-strings. This expression has been conjectured
\refs{\rfRusso} to be some sort of approximation to the exact four-graviton amplitude
of the type IIB theory. It does indeed reproduce  the first few of
the known coefficients in the low energy expansion: by definition, it contains
the exact tree-level amplitude and it also contains the correct ratio of
the $E_{3/2}\, \calR^4$ term and the  $E_{5/ 2}\,
{\alpha'}^2 \,(s^2+t^2+u^2)\, \calR^4$ term.
However, it  produces a value for the coefficient
  of the  one-loop part of the   ${\alpha'}^3 \,(s^3+t^3+u^3)\, \calR^4$
interaction that is twice the value of  $c$ in \summres.
 It  is not surprising that the naive
modular invariant conjecture of \refs{\rfRusso}  fails since there is
no obvious  sense in which it can approximate
 the exact amplitude.  After all it purports to describe  an infinite number of
 highly unstable non-BPS states in a non-perturbative manner
 but lacks all of
the (massless and massive) threshold cuts that are required by unitarity.

\vskip 0.5cm
{\it Acknowledgements:}
 P.V. is grateful to PPARC for financial support and
 the Service de Physique th{\'e}orique de Saclay for hospitality.
  Both authors are
  also grateful to the organizers of the Amsterdam Workshop (July 5-16
 1999) and to the Theory Division at CERN where this work was completed.


\appendix{A}{Massless normal thresholds}

The thresholds that arise from massless on-shell intermediate states come from the
region of integration over  near the  boundary of  moduli space at  which the
toroidal world-sheet pinches   in such a manner that  the four
vertex operators are separated into two bunches.  At this
degeneration point the world-sheet is the product of the
two tree-level world-sheets that enter in the right-hand side of
\discone.

In order to extract these thresholds from the expression \eF\ for
the loop amplitude it is very useful to change the definition of
the moduli from $\nu^{(r)}$ and $\tau$ to $\eta^{(r)}$ by
defining
\eqn\etadef{\eqalign{
\nu^{(1)}=\eta^{(1)},&
 \qquad \nu^{(2)}=\eta^{(1)}+\eta^{(2)}
,\qquad \nu^{(3)}=\eta^{(1)}+\eta^{(2)}+\eta^{(3)}, \cr
& \nu^{(4)}=\tau=\eta^{(1)}+\eta^{(2)}+\eta^{(3)}+\eta^{(4)},\cr}}
where we have used the conformal invariance of the loop amplitude
to fix $\nu^{(4)} =\tau$.  The $\eta$ variables are the ones that
arise naturally in the operator  construction of the loop amplitude as a
trace over a string tree.   In such a construction the
propagator describing each leg of the loop is written as
\eqn\eProp{
\Delta_i = {\alpha'\over 2\pi} \int_{|z|<1} {dzd\bar z\over |z|^2} \; z^{L_0}
{\bar z}^{\bar L_0},
}
where $z=e^{2\pi i \eta}$.

The   degeneration limit of relevance to the $s$-channel thresholds is
the one in which  $\eta^{(1)}_2 \to \infty$ and
$\eta^{(3)}_2\to \infty$, which puts the two s-channel propagators in the loop
 on shell.
This corresponds to the region
of integration $\C|T_{st}$:
\eqn\regdef{  \nu^{(1)}_2\leq \nu^{(2)}_2 \leq \nu^{(3)}_2 \leq
\nu^{(4)}_2=\tau_2}
with  $\tau_2\to \infty$.
In this limit we may substitute the asymptotic values,
\eqn\delttu{\Delta_t\sim \tilde\Delta_t=\hat
\C|P^\infty(\nu^{(14)})+\hat \C|P^\infty(\nu^{(23)}),\qquad
\Delta_u \sim \tilde\Delta_u=\hat \C|P^\infty(\nu^{(13)})+\hat
\C|P^\infty(\nu^{(24)}),}
and
\eqn\edeltass{
\eqalign{ \Delta_s& \sim  {\pi \left(\nu^{(12)}_2\right)^2\over 2\tau_2}-
  {1\over4} \ln\left|\sin(\pi \nu^{(12)})\over \pi \right|^2+ {\pi
    \left(\nu^{(34)}_2\right)^2\over 2\tau_2}-
  {1\over4} \ln\left|\sin(\pi \nu^{(34)})\over \pi \right|^2\cr
&=  \tilde\Delta_s+    \delta_s,
}}
where
\eqn\sdelsdef{\tilde\Delta_s=\hat \C|P^\infty(\nu^{(12)})+\hat
\C|P^\infty(\nu^{(34)}),}
 and
\eqn\sdeltm{  \delta_s  = \sum_{m\neq 0} {1\over 4|m|} \left(e^{2i\pi
    (m \nu_1^{(12)}+i |m| \nu_2^{(12)})} +e^{2i\pi
    (m \nu_1^{(34)}+i |m| \nu_2^{(34)})} \right).}
The sum over $m$ in \edeltass\ and \sdeltm\
 gives the effect of the  massive string states
that propagate between the  vertices for the particles $1$ and $2$
or the vertices for the  particles $3$ and $4$,  i.e.
in the legs of the loop that are not degenerating.   These terms
are the ones that
give rise to the stringy corrections to the low-energy field
theory thresholds.

The contribution to the one-loop amplitude in the $\C|T_{st}$
region  can be rewritten as
\eqn\eLoopdeg{
I_{\calT_{st}} = \int_{\calR_\L}^\infty {d^2\tau\over
\tau_2^2} \int_{\C|T_{st}} \prod_{i=1}^3 {d^2\nu^{(i)}\over
\tau_2} \exp\left(\alpha' s (\tilde\Delta_s - \tilde\Delta_u)
+\alpha' t (\tilde\Delta_t-\tilde\Delta_u)  \right) \
\exp\left(\alpha's \delta_s\right) }
The $\alpha'$ expansion is obtained by expanding the last exponential in
powers of $\delta_s$. The leading term reproduces the field theory $s$-channel
threshold  given by the first term in~\qresult.  The next contribution, linear
in $\delta_s$, vanishes  due to the integration over $\nu_1^{(2)}$ or
$\nu_1^{(4)}$. The next   term  has a factor of $({\alpha's}\,  \delta)^2$ and
gives a non-zero contribution to the logarithmic behaviour at order
$(\alpha's)^4$. After a little algebra (and adding the contributions of the
$\calT_{tu}$ and $\calT_{us}$ domains) this gives the threshold  contribution
\eqn\eSfourCut{\eqalign{
I_{nonan\;2}&(s,t, -s-t)= \sum_{m\neq 0}{(\alpha's)^2\over 32m^2}
\!\int_\L^\infty\!\! {d \tau_2\over \tau_2^2}\!
 \int_{\C|T_{st}}\! \prod_{i=1}^3
d\omega_i e^{\alpha'\pi \tau_2 Q(s,t)}
 \cr &   \left(e^{-4\pi m
\tau_2(\omega_2-\omega_1)}+e^{-4\pi m \tau_2 (1-\omega_3)}
\right) + tu\ {\rm term} + us\ {\rm term}
,\cr}}
where the integration variables $\omega_i$ are defined in \rescnu.
This integral is complicated but for the special case $t=0$ it reduces to
the simple expression
\eqn\eSfourRes{
I_{nonan\;2}(s,0,-s)= {4\over\pi^3} \zeta(3)\zeta(4)\;
\left(\alpha's\over 4\right)^4  \left(\ln\left( {\alpha' s\over 4\L}
  \right)+ \ln\left(-  {\alpha' s\over 4\L} \right)\right).
}

\listrefs

\bye